# Performance testing of lead free primers: blast waves, velocity variations, and environmental testing

Authors:   Elya Courtney[1]
           Amy Courtney[1]
           Peter David Summer[2]
           Michael Courtney[1]

Results are presented for lead free primers based on diazodinitrophenol (DDNP) compared with tests on lead styphnate based primers. First, barrel friction measurements in 5.56 mm NATO are presented. Second, shot to shot variations in blast waves are presented as determined by detonating primers in a 7.62x51mm rifle chamber with a firing pin, but without any powder or bullet loaded and measuring the blast wave at the muzzle with a high speed pressure transducer. Third, variations in primer blast waves, muzzle velocities, and ignition delay are presented after environmental conditioning (150 days) for two lead based and two DDNP based primers under cold and dry (-25° C, 0% relative humidity), ambient (20° C, 50% relative humidity), and hot & humid (50° C, 100% relative humidity) conditions in 5.56 mm NATO. Taken together, these results indicate that DDNP based primers are not sufficiently reliable for service use.

## INTRODUCTION

Primer technology has a cyclical history with notable instances of new primer chemistry being introduced due to environmental or maintenance concerns with several decades passing before the new chemistry became reliable. In 2010, the Office of the Product Manager for Maneuver Ammunition Systems (USA) projected that lead free primer formulations for use in the U.S. military would be evaluated and candidates selected in 2011, and that ammunition with green primers would be at full production by the end of 2012.[1] These projections proved to be overly optimistic.

Primer development at the turn of the 20$^{th}$ century was driven by the need for a non-corrosive formulation. In the following years, changes in primers used by the military were necessary due to lack of shelf stability, which led to misfires. This was a reason the U.S. military moved from mercury fulminate based primers prior to WWI to a formulation based on potassium chlorate, antimony trisulfide, and sulfur. However, this formulation was associated with misfires and corrosion, forcing another change.[2]

Removing lead from ammunition due to environmental and health concerns has been a recent focus in ammunition development. Since the U.S. issued the new M855A1 load with lead-free bullets, lead-based rifle primers are a primary source of lead remaining in duty ammunition.[3] Diazodinitrophenol (2-diazo 4,6 dinitrophenol, abbreviated as DDNP and also referred to as diazole and dinol) has long been considered a promising candidate to replace lead styphnate in centerfire priming compounds. An Air Force study [3] showed that transitioning to training ammunition with lead-free bullets and primers can reduce instructor exposure to lead by 70% in indoor ranges and 41% in outdoor ranges.

---

1 BTG Research, Baton Rouge, Louisiana
michael_courtney@alum.mit.edu
2 U.S. Air Force Academy, 2354 Fairchild Drive, USAF Academy, CO 80840

Diazo compounds were discovered by Peter Griess in 1858.[4] Over the next 50 years, the chemistry of diazo compounds and their applications were further investigated. In the 1980s, interest increased in using DDNP in non-toxic priming compounds, and since 1985 more than a dozen patents have been awarded for small arms applications. Between 2000 and 2011, several American ammunition manufacturers offered product lines with lead-free primers, and DDNP based primers were being manufactured on a commercial scale at Murom in Russia.[2]

This paper presents results for three experiments on lead free primers based on diazodinitophenol (DDNP) comparing results with identical tests on lead styphnate based primers. First, barrel friction measurements in 5.56 mm NATO are presented for lead based and lead free primers. Second, shot to shot variations in blast waves are presented as determined by detonating primers in a 7.62x51mm rifle chamber with a firing pin, but without any powder or bullet loaded and measuring the blast wave at the muzzle with a high speed pressure transducer. DDNP based primers have significantly larger variations in performance than lead based primers. Third, variations in primer blast waves, muzzle velocities, and ignition delay are presented after environmental conditioning (150 days) for two lead based and two DDNP based primers under cold and dry, ambient, and hot & humid conditions in 5.56 mm NATO. The DDNP based primers showed significant ignition delays and greater velocity variations in all treatment groups along with 90-100% misfire rates in the hot & humid treatment group.

**FRICTION EFFECTS IN 5.56x45mm NATO**

A new method of measuring barrel friction has been recently developed to determine average barrel friction over the length of a rifle barrel at ballistic velocities.[5] This method has been used to test purported friction reducing effects of various coatings, with the findings that most coatings do not offer significant reductions in barrel friction.[6] The original study [5] mentioned preliminary data showing an increase in barrel friction associated with lead free primers based on diazodintrophenol (DDNP).

Increased barrel friction can result in higher operating pressures and rob projectiles of kinetic energy. Lower velocity projectiles have more drop and wind drift with distance. Lead styphnate based primers may also contribute to the lubrication and reliable operation of the AR based M-16 and M-4 rifles fielded by various branches of the military. If lead based fouling combines with the applied lubricants in a synergistic manner to maintain feeding and functioning, then a change to a lead free primer may inadvertantly reduce system reliability.

---

3   This is an oversimplification. Most sniper and counter sniper loads still use match style jacketed lead bullets. Available lead free bullet technologies are inferior to jacketed lead bullets in several key areas essential to long range effectiveness: accuracy, barrel friction, aerodynamic drag, and wind drift. For example, there is no lead free bullet offering anywhere near the performance of the 220 grain Sierra MatchKing in the .300 Winchester Magnum Marine sniper load. Lead free bullets are inherent compromises for applications beyond 300 m.

This experiment evaluated dependence of barrel friction on primer type in 5.56 mm NATO for available small primers. Most manufacturers of DDNP based primers only sell these primers as components in loaded lead free ammunition. To the authors' knowledge, the only exceptions are the Russian made primers from the factory in Murom, which have been imported to the U.S. and marketed under the PMC, Wolf, and Tula brands. The authors contacted ATK, Winchester, and Remington to request component DDNP based primers for testing; however, none of these US manufacturers provided component primers for testing. Knowing that ATK lead free primers are used in Air Force training ammunition and are also being offered by ATK for field use by the US military, the authors acquired DDNP based primers from ATK by purchasing fully loaded lead free ammunition.

The method for determining barrel friction has been described previously. [5,6] A high quality match grade bullet with a thin, precision jacket, soft lead core, and tight weight tolerance was chosen for this experiment rather than one of the military projectiles such as the M193 (55 grain full metal jacket) or the M855 (62 grain penetrator core). These military projectiles show larger variations in hardness, dimensions, and weight tolerances, likely leading to greater variations in barrel friction and muzzle velocity and likely introducing confounding factors, when the experimental goal was to isolate the influence of primer type on barrel friction.

Twenty five bullets were loaded for each combination of primer type and powder charge. When the average energy for five shots was graphed vs. powder charge, the resulting graph illustrated a strong linear relationship with a coefficient of determination consistently above 0.995. The best-fit slope is the additional energy obtained for each additional grain of powder. The vertical intercept is negative and represents the mechanical work necessary to overcome resistive forces in the barrel. High linear correlation gives confidence that the muzzle energy is truly a linear function of powder charge for the choice of bullet and powder so that extrapolating back to the vertical intercept to determine the friction is valid. The uncertainty in energy at each powder charge is small (< 1%) due to bullet choice and careful barrel cleaning and reloading procedures.

ATK DDNP based primers were acquired as components in loaded ammunition (Federal Cartridge Company part number BC556LTOM1). After the bullet was pulled with a collet type puller, the factory ball powder was removed from the case, and the cases were loaded with experimental charges of Alliant Blue Dot powder and a 62 grain Berger Flat Base bullet was carefully seated.

The energy lost to friction was 376 ft lbs (+/- 35 ft lbs) when using the Russian made (Murom) DDNP based primer, which was significantly greater than the 330 ft lbs (+/- 2 ft lbs) lost to friction using a lead based primer (ATK Fed 205m primer). The American made DDNP based primer (ATK) produced a measured 322 ft lbs (+/- 40 ft lbs) lost to friction. The large uncertainty in the friction determination with the ATK DDNP based primer was caused by velocity variations and made its friction statistically indistinguishable from either the Murom lead free primer or the Fed 205m lead based primer.

*Figure 1: Largest and smallest blast pressure waves for designated primers. Arrows denote the range of peak blast pressures for a sample size of 10. (Waveforms for different models are offset in time to facilitate comparison.)*

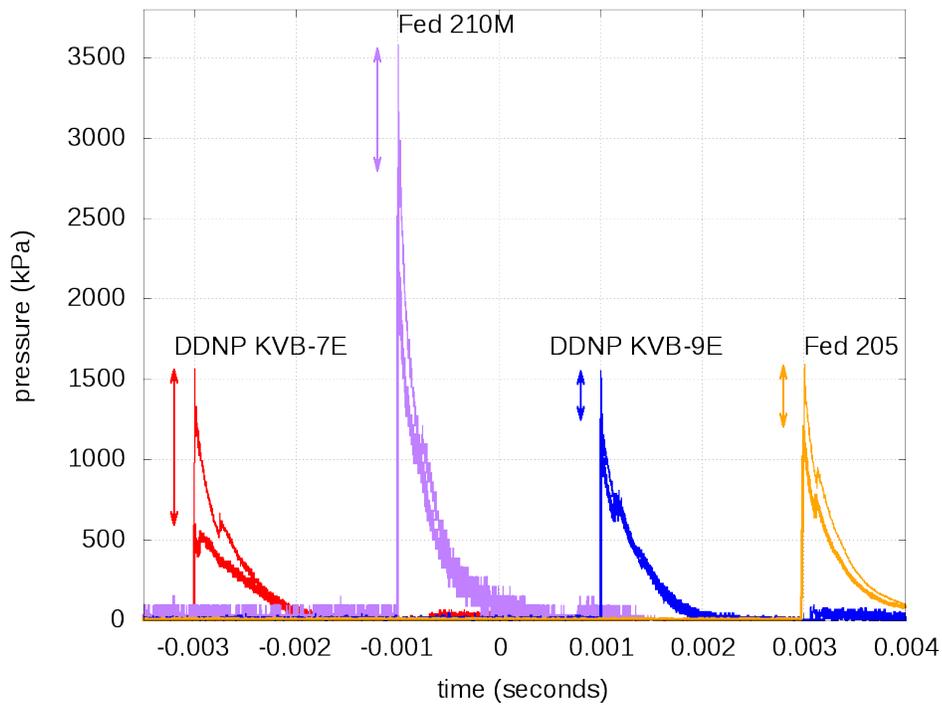

| Primer | Size | Peak Pressure (kPa) | SD (kPa) | SD (%) |
|---|---|---|---|---|
| Fed 210M | Large | 2908 | 223 | 7.7% |
| Fed 215M | Large | 3811 | 192 | 5.0% |
| CCI 200 | Large | 2561 | 270 | 10.7% |
| CCI 250 | Large | 3587 | 404 | 11.3% |
| **DDNP KVB-7E** | **Large** | **1186** | **296** | **25.0%** |
| Rem 7 ½ | Small | 2303 | 186 | 8.1% |
| Fed 205 | Small | 1469 | 103 | 7.1% |
| CCI 450 | Small | 1602 | 104 | 6.5% |
| Fed 205M | Small | 1434 | 103 | 7.2% |
| **DDNP KVB-9E** | **Small** | **1331** | **109** | **8.2%** |

*Table 1: Peak pressure averages and standard deviations from the mean (SD).*

**BLAST PRESSURE VARIATIONS IN 7.62x51 NATO**

Full assessment of primer performance requires testing fully loaded ammunition. However, a method to test primer performance independently of confounding effects of powders, bullets, neck tension, case capacity, and bore friction has the advantage of more direct comparison. The method here measured blast pressure produced by impact detonation of a primer loaded in a cartridge case without any bullet or powder.

A high-speed pressure transducer (PCB 102B or PCB 102B15) was placed at the muzzle of a 7.62x51mm NATO rifle barrel with no separation between the end of the barrel and pressure transducer. The voltage waveform was converted to pressure using the calibration certificate provided by the sensor manufacturer. Ten trials each of eight common models of lead styphnate based primers were tested along with two DDNP based primers, a large rifle primer (model KVB-7E) and a small primer (model KVB-9E) manufactured in Russia at the "Murom apparatus producing plant."

Blast pressure waveforms of four primer types are shown in Figure 1. Table 1 shows mean peak pressures along with standard deviations for primers studied. Except for the DDNP based large rifle primer, large rifle primers produce stronger blast waves than small primers, and "magnum" rifle primers (Fed 215M, CCI 250) produce stronger blast waves than non-magnum primers of the same size. Different primer types have significant differences in their standard deviations, and it is notable that so-called "Match" primers are not always more consistent than non-match primers. In each group (large and small), the standard deviation of the DDNP based primer is the largest percentage of its mean value. For the large rifle primers, the standard deviation of the DDNP based primer (25%) is more than twice the standard deviation of any lead styphnate based primer that was tested.

Minimal field testing was also conducted comparing 10 shots with the DDNP based rifle primer with 10 shots of the lead styphnate based Fed 210M in each of two otherwise identical loads: 1) a 30-06 load using 51.0 grains of H414 (a ball powder) in Remington brass with a 220 grain Sierra MatchKing bullet and 2) a 7.62x51mm NATO load using 46.0 grains of Varget (an extruded powder) using Remington brass with a Berger 155.5 grain Fullbore boat tail bullet. Both tests were conducted with Remington 700 bolt action rifles. There was a perceptible delay between firing-pin strike and ignition in 15 of 19 shots with the DDNP based primers (and one misfire); in contrast, there were no misfires or perceptible delays in ignition with the lead styphnate based primer.

**EFFECTS OF ENVIRONMENTAL CONDITIONING**

This experiment compared performance of lead based and lead free primers after conditioning for 150 days in cold, ambient, and hot & humid environments. Performance measures included muzzle velocity variation, ignition delay, and primer peak blast pressure variation. Weak and/or unpredictable ammunition may result in the loss of life in duty applications. The military has specifications stating that there may not be more than a four millisecond delay between firing pin strike and powder ignition. [7]

Four types of primers were tested. The first primer type was a conventional non-magnum, lead styphnate, small rifle primer made by Cascade Cartridge Inc., a subsidiary of Alliant Techsystems (ATK). This primer will be denoted as CCI. The second primer type was a CCI Magnum lead styphnate based primer, denoted as MCCI. The third primer type was a DDNP based lead free small primer from ATK, denoted as ATK-LF. This is the same model and lot number as the ATK lead free primer used in the friction study above, and primed cases were obtained from loaded ammunition in an identical manner. The fourth primer was the Murom KVB-9E DDNP based primer, referred to as KVB-LF.

The primers (20 of each primer type per each environmental condition, 240 primers total) underwent environmental conditioning in foreseeable environments. The three environmental conditions were: **cold** – deep freeze of -25°C with 0% relative humidity, **ambient** – ambient indoor conditions with 40%-60% relative humidity and temperatures averaging 20°C, **hot & humid** – chamber maintained at 50°C with 100% relative humidity. The exposure time was 150 days. Temperature was monitored and recorded every 12 hours with thermocouple based probes.

Loads were identical other than the primer with Lake City brass, 55 grain FMJ bullets, and 25 grains of spherical powder (Winchester 748). After completing environmental conditioning of primers, ten cartridges were loaded for each primer type and treatment group, 120 loads total. Velocities were measured with a CED Millenium M2 optical chronograph with infrared skyscreens (0.3% accuracy). Cartridges were fired from a Remington 700.

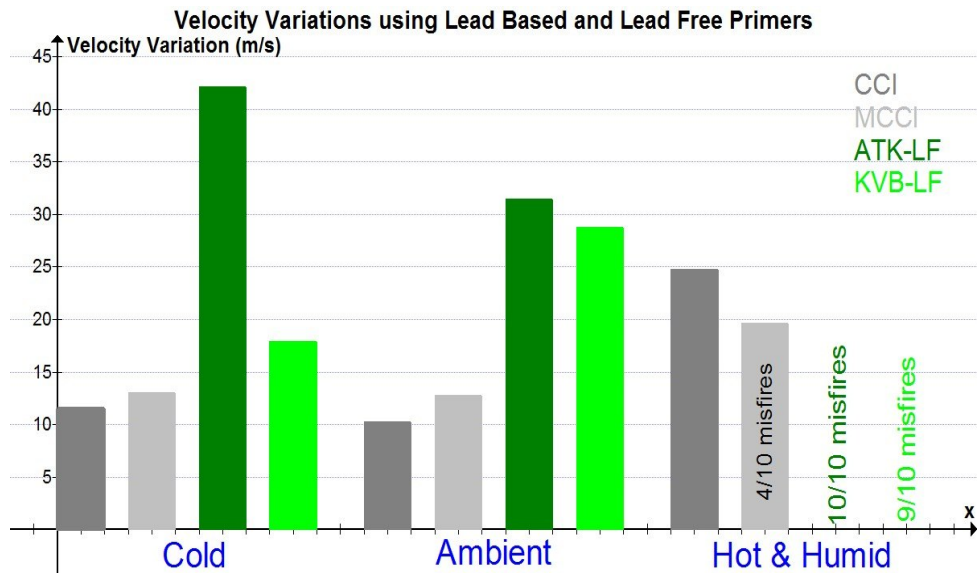

*Figure 2: Standard deviation in muzzle velocity. Note lead styphnate based primers have smaller velocity variations and fewer misfires than DDNP based primers.*

In muzzle velocity tests, nearly all (19 of 20) DDNP based primers misfired in the hot & humid treatment group. As shown in Figure 2, DDNP based primers had larger velocity variations in the cold and ambient treatment groups. MCCI and CCI primers had comparable velocity variations in cold and ambient treatment groups. MCCI primers had 4/10 misfires in the hot & humid treatment group. Overall, muzzle velocity results showed lead styphnate based primers have less variation in muzzle velocity than DDNP based primers.

Ignition delay was determined with a microphone placed 45 cm from the chamber to the side of the rifle to record the sounds of firing pin strike and powder charge ignition. After the trigger was pulled, the sound waveform recorded key events in the firing sequence at a rate of 100,000 samples per second. Key events included the movement of the firing pin, the firing pin striking the primer, and ignition of the powder. Recording was done simultaneously with muzzle velocity testing.

Ignition delay data, summarized in Figure 4, shows that for each treatment group, lead styphnate based primers outperformed DDNP based primers by having a shorter ignition delay. For the CCI primers in hot & humid conditions, the average and standard deviation are noticeably larger. For lead styphnate based primers, CCI outperformed MCCI, but lead styphnate based primers performed better than DDNP based primers for all conditions. For lead free primers, there was a delay in almost every trial that was not a misfire; furthermore, these delays were inconsistent, showing additional reason to doubt performance reliability of DDNP based primers. Hot & humid conditioning was by far the most detrimental to performance for all primers tested. DDNP based primers failed to ignite the powder (with one exception), and the lead styphnate based primers had ignition delays.

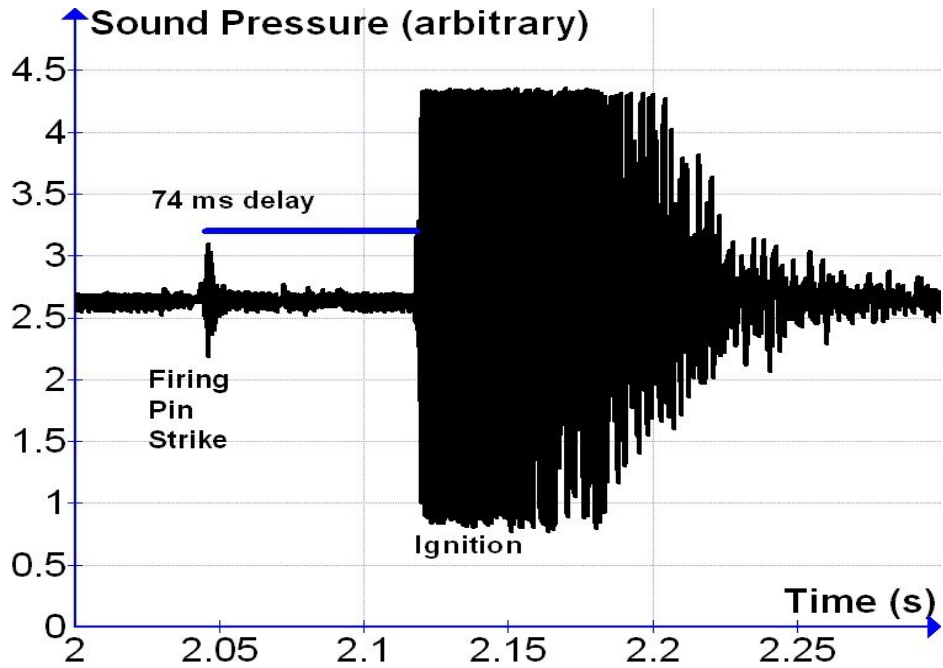

*Figure 3: Sound waveform for first trial of the ATK-LF primer from the cold treatment group. Analysis of the sound waveform shows a 74 ms delay between the firing pin strike and powder ignition.*

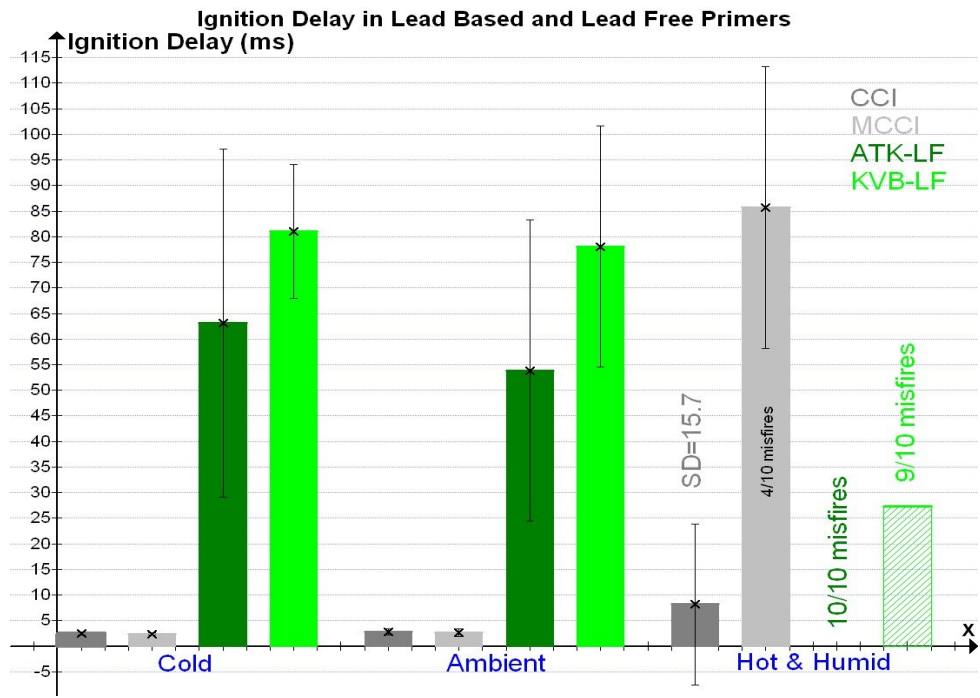

*Figure 4: Ignition delays for each primer type and treatment group. Error bars are the standard error of the mean (SEM). Large error bars reflect inconsistent performance.*

To measure the blast waves produced by each primer, primed 5.56mm NATO cartridge cases (no bullet or powder) were placed in the chamber of the test rifle, and a PCB Piezotronics (113B24) high speed pressure transducer was placed directly inside the muzzle of the rifle barrel. There were 10 trials each of each primer type and environmental treatment group (120 trials total).

As shown in Figure 5, blast waves produced by primer detonations produced results that were consistent with the results from the muzzle velocity tests. Hot & humid conditioning degraded performance of all primer types, especially the DDNP based primer types. Peak blast pressures produced by the ATK-LF primers was at least three times higher than peak blast pressures produced by the KVB-LF primers for cold and ambient conditions. The CCI and MCCI primers had comparable performance for the cold and ambient conditioning, but CCI primers exposed to heat and humidity produced a higher peak blast pressure than MCCI primers exposed to the same.

ATK-LF primers had the highest peak blast pressures, but failed to detonate when exposed to heat and humidity. KVB-LF primers had the lowest peak blast pressures. When the KVB-LF primers exposed to heat and humidity were tested, only one primer detonated and it had the lowest peak pressure. DDNP based primers are not suitable for field or duty use due to ignition delays and/or misfires in all treatment groups. Lead based primers performed better in all treatment groups.

After hot & humid conditioning, MCCI and KVB-LF primers had crystal residue outside the primer cup. This suggests an ionic compound (likely an oxidizer) dissolved in moisture and crystallized outside the cup when the moisture evaporated. The number of misfires of primers exposed to hot & humid conditions was the same for the blast wave testing and the muzzle velocity/ignition delay testing.

The CCI and MCCI primers had similar performance in the cold and ambient conditions, but the CCI primers clearly handled hot & humid conditioning better than the MCCI primers. The ATK-LF primers had higher peak pressure and shorter ignition delay, but the KVB-LF primers had less velocity variation. Neither primer type responded well to the hot & humid conditioning, but the KVB-LF had less variation between the cold and ambient conditioning results than the ATK-LF had for the same.

The manufacturer (ATK) of one of the DDNP based primers tested included a small clause on the box of its ammunition; stating "Extended storage at an elevated temperature may degrade performance and result in misfires." The CCI and MCCI primers had similar performance in the cold and ambient conditions, but the CCI primers clearly handled the hot & humid conditioning better than the MCCI primers.

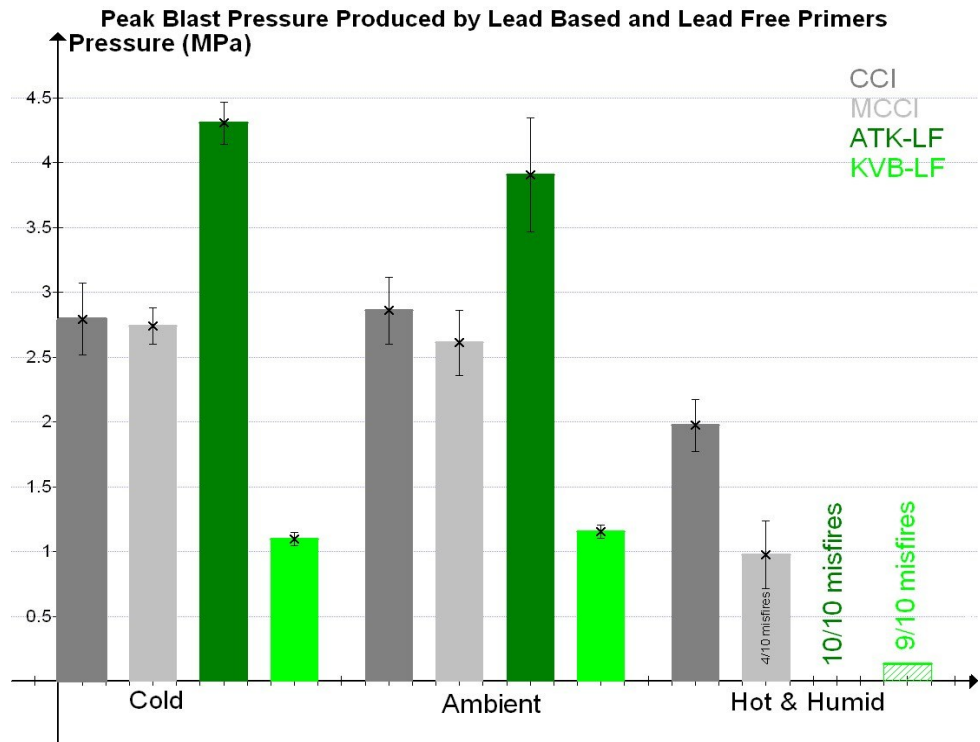

*Figure 5: Peak blast pressure produced for each primer type and treatment group.*

**DISCUSSION**

Efforts to date seemed to have focused on DDNP based primers as the most promising replacements for lead based compounds in small arms primers. No viable commercial or military product has emerged that is suitable for duty use. There are hints in patent applications and other publications that ATK and ARL are pursuing alternate technologies including micronized aluminum and red phosphorus as potential lead free priming candidates.

Several new methods for quantifying important aspects of primer performance have been demonstrated in this paper. These methods are particularly useful because they can be employed in any rifle of the appropriate caliber without the need to modify the rifle to accommodate specialized instrumentation. The method for measuring barrel friction depends on the relationship between muzzle energy (determined with a chronograph) and powder charge (measured with a precision scale when ammunition is loaded). This allows barrel friction to be determined in any rifle without modification. The method for measuring primer blast waves employs a high speed pressure transducer at the rifle muzzle rather than a customized chamber and firing pin. Ignition delay is measured with a microphone rather than an instrumented receiver, both allowing ignition delay to be determined in any service rifle, as well as measuring ignition delay simultaneously with velocity variations in any given rifle.

## ACKNOWLEDGEMENTS

We thank Louisiana Shooters Unlimited for use of their range in Lake Charles, Louisiana, and Dragonman's Range for use of their range in Colorado Springs, Colorado. This work was funded by the US Air Force Academy and BTG Research, who played no role in manuscript preparation other than employment of some authors.